\documentclass{article}
\usepackage{graphicx} 
\usepackage{float}
\usepackage{tabularray}
\usepackage{booktabs}
\usepackage{breqn}
\usepackage{rotating}
\usepackage{hyperref}
\usepackage{setspace}
\usepackage{kotex, amsfonts, amsmath, amssymb, amsthm, bbm}
\usepackage{natbib}
\usepackage{xcolor}
\usepackage{siunitx}
\usepackage[
  top    = 1cm,   
  bottom = 1cm,
  left   = 2cm,
  right  = 2cm
]{geometry}

\font\myfont=cmr12 at 20pt
\newcommand{\tit}{ \textbf{\large Sticky Homelessness (Working Paper - December 31, 2025 Version) 
}\thanks{I would like to thank Professor *** for his invaluable guidance on this project. Many thanks to HUD staff and ***, Ph.D of BLS for their generous help in understanding and accessing resources. I would also like to thank U.C. Berkeley Ph.D. candidates *** and *** for being awesome mentors throughout my time in Berkeley. Lastly, I want to thank *** and the Night on the Streets Catholic Worker (NOSCW) organizers for the opportunity to volunteer and learn firsthand about the homelessness crisis. }
}
\title{\myfont \tit } 
\author{Richard Yun\thanks{Seoul National University}}
\date{September 2025}

\begin{document}

\maketitle

\begin{figure}[H]
  \centering
  \href{https://homelessnessatlas.shinyapps.io/The_Test/}%
       {\includegraphics[width=0.7\textwidth]{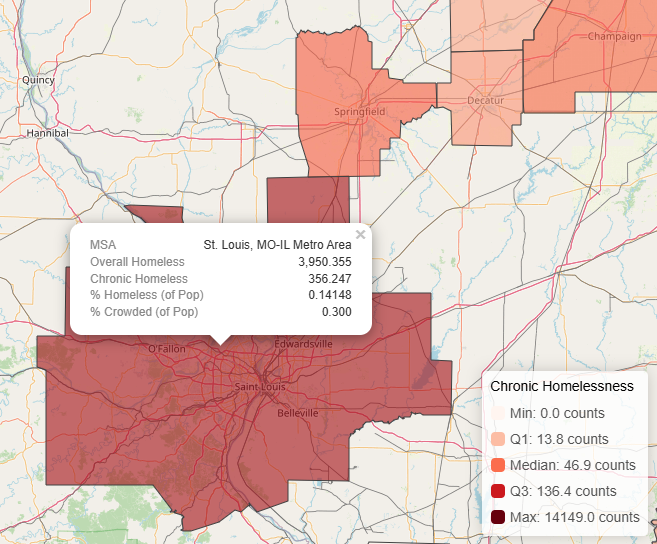}}

  \caption{%
    \href{https://homelessnessatlas.shinyapps.io/The_Test/}{%
      \mbox{\colorbox{yellow!30}{\bfseries\textcolor{blue!70!black}{Click}}~on~this~figure}%
    } to view the%
    \href{https://homelessnessatlas.shinyapps.io/The_Test/}{%
      \mbox{\colorbox{yellow!30}{\bfseries\textcolor{blue!70!black}{Metro~Homelessness~Atlas}}}%
    }.%
  }
  \label{fig:metro_atlas}
\end{figure}

\begin{abstract}

Homelessness in American cities is becoming an ever more prominent issue, but its causes remain contested, ranging from mental health and substance abuse to housing affordability and local labor markets. To shed light on this issue, I construct a novel MSA-level national panel of homelessness counts using data from the U.S. Department of Housing and Urban Development. Using a long-differencing regression specification with the changes in rent entered in piecewise linear form, I find that rent increases predict large increases in homelessness rates, but decreases have little to no effect. The same conclusions are reached when I use a quasi-differencing moment condition, assuming a multiplicative mean specification. Then, I propose a theoretical model of the low-end housing market that explains the asymmetry I find in the data. Finally, I outline an IV strategy that instruments rent changes with a Bartik instrument of predicted employment growth interacted with local housing-supply elasticities. My findings suggest that homelessness \textbf{is} a housing problem; however, because the response is sticky downward, effective policy must complement housing-market interventions with measures that address barriers faced by people experiencing homelessness.

\end{abstract}


\newpage

\section{Main Graph and a 1-Page Summary}

\subsection{Key Findings}

\begin{enumerate}
  \item \textbf{Downside stickiness in homelessness.} A 1\% \emph{increase} in MSA median rent is associated with a 1.7–2.1\% increase in the chronic homelessness rate. In contrast, a \emph{decrease} in rent has an effect that is statistically indistinguishable from zero and economically small.

  \item \textbf{Robust across estimators.} The asymmetric response appears in both a first-differenced OLS specification and in a quasi-differenced specification estimated by the method of moments.

  \item \textbf{Model} A simple model of the low-quality housing market explains the asymmetry: past homelessness depresses future earning capacity, so agents benefit less from rent declines than they are harmed by rent increases.

  \item \textbf{Causal Estimates (forthcoming)} I outline an IV strategy that instruments rent changes with a Bartik (shift–share) instrument of predicted employment growth interacted with local housing-supply elasticities and observed heterogeneities in supply constraints. The design leverages the fact that labor-demand shocks move local housing costs, with larger price responses where housing supply is inelastic.
\end{enumerate}

\begin{figure}[H]
  \centering
  \includegraphics[width=0.62\textwidth]{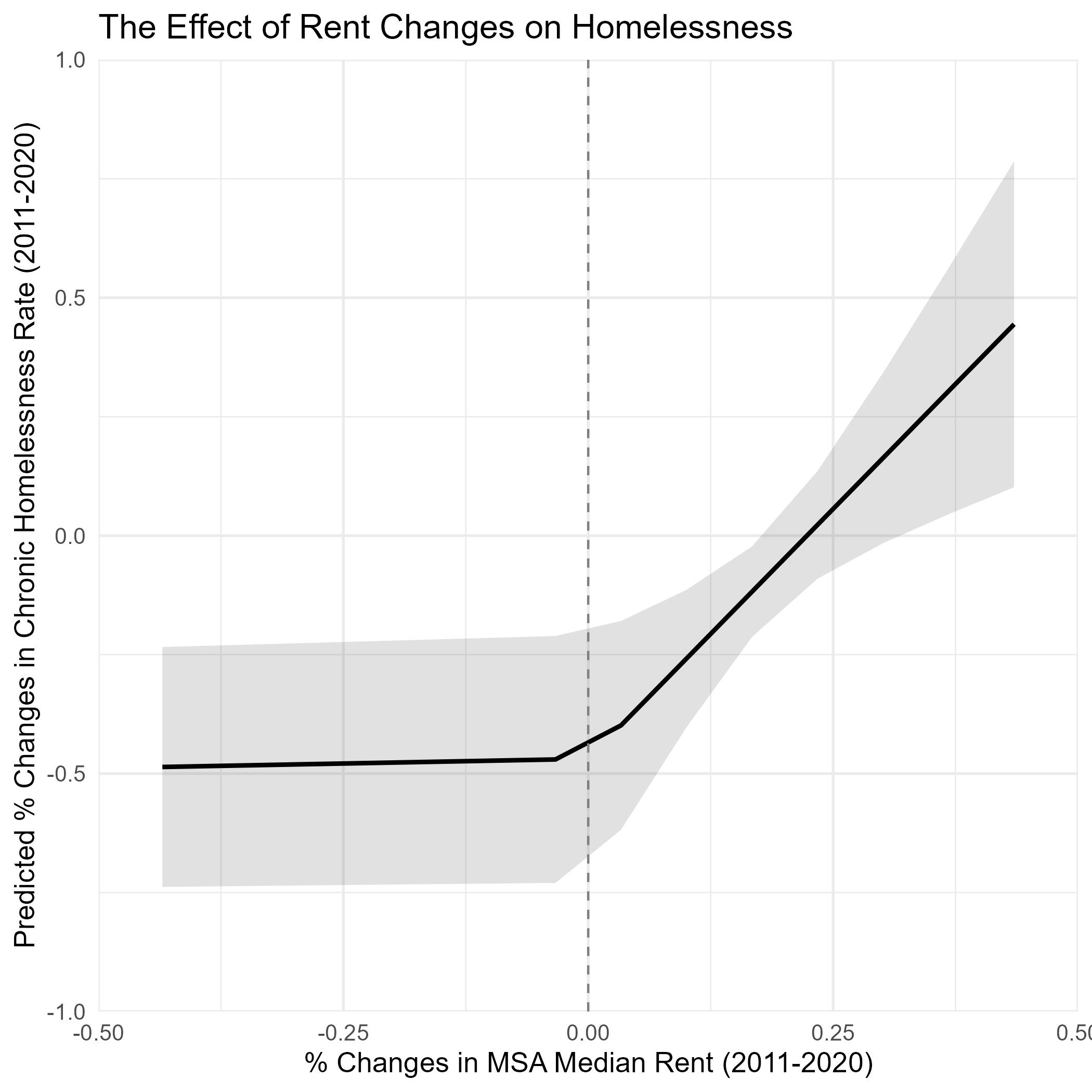}
  \caption{Margins Plot for  Table~3, Col (3)\\ Shaded area: 95\% pointwise CI}
  \label{fig:margins}
\end{figure}

\subsection{Contributions to the Literature}

\begin{enumerate}
  \item \textbf{Data.} Assemble and analyze a new \emph{national panel} of homelessness counts at the MSA level.
  \item \textbf{Measurement.} Use MSA rents directly from the ACS, avoiding imputation at the CoC level and improving precision.
  \item \textbf{Results.} Document an \emph{asymmetric} elasticity of chronic homelessness rates with respect to rent. (work in progress develops causal identification via IV)
\end{enumerate}

\newpage

\section{Data}

\textbf{\large Continuum of Cares and their Limitations}
\normalsize\\
\href{https://www.hud.gov/hud-partners/community-coc}{Continuum of Cares (CoC)} are local planning bodies responsible for coordinating the full range of homelessness services in a geographic area, which may cover a city, county, metropolitan area, or an entire state. During the last week in January of each year, CoCs conduct an unduplicated count of both sheltered and unsheltered people experiencing homelessness within their jurisdiction, called a 'Point-In-Time' count. The PIT data is released every year at the CoC level and is the basis of the HUD's Annual Homelessness Assessment Report (AHAR) to Congress.

Several studies have used HUD’s Point-in-Time (PIT) counts to infer structural drivers of homelessness, but these efforts face two key limitations. First, CoC-level measures of housing markets and local economic conditions are not reported. As a result, work that treats the CoC as the unit of analysis must impute CoC variables from other geographies. For example, Byrne, et. al (2012) construct CoC-level median rents by taking the population-weighted average of the median rents of all counties whose centroid falls under a CoC boundary. Conrith (2015) follows the same method.

This imputation strategy has at least two serious drawbacks. First, the population-weighted median rent is only a proxy for the true CoC median and can be highly noisy. The noise could be substantial in large CoCs like \href{https://files.hudexchange.info/reports/published/CoC_Dash_CoC_TX-607-2023_TX_2023.pdf}{TX-607}
(Texas Balance of State CoC), as it contains 216 counties in its jurisdiction. TX-607 highlights a more fundamental issue with analysis at the CoC level, which is that some CoCs are simply too big. It is unrealistic to think that one value can capture the rental market conditions of an area that covers more than half of Texas.

Second, CoC boundaries are not stable over time. Frequent mergers and splits make it difficult to construct a consistent panel. In 2011, for instance, CA-601 merged with and absorbed CA-610 and reported 1,053 more homeless individuals than in 2010 (about 12\% growth). Without harmonizing boundaries or adjusting counts, it is impossible to separate the jump caused by the merger from genuine changes in conditions within the original CA-601 area. Without some form of correction, it is impossible to determine how much of this large increase in homelessness (12\%) should be attributed to the merger itself and to changes in the economic and societal conditions in the initial jurisdiction of CA-601. 

To address boundary instability, Hanratty (2015) drops all CoCs that changed boundaries between 2007 and 2014. This removes 63 of the 462 CoCs observed in 2007—a sizable share—and drops major metropolitan areas such as San Diego, which was covered by CA-610 (San Diego County CoC) and later CA-601 (San Diego City and County CoC). Corinth (2015) takes a different approach by combining CoCs in years before mergers to maintain geographic consistency. However, in doing so, he is making geographic boundaries larger, worsening the 'noisy rent' problem illustrated above. This could be the reason that he does not find statistically significant coefficients for median rent in any of his model specifications. 

To circumvent these limitations surrounding CoC-level analysis, I construct a new MSA-level panel of homelessness counts using modern geocomputation techniques.\footnote{I share credit with Cynthia Shi, a UC Berkeley Economics and Statistics major, for our dataset \textit{Metro Homelessness Atlas}}. Working at the MSA level has two key advantages. First, it allows direct use of ACS median rents of MSAs published by the Census Bureau—no imputation or aggregation from counties—providing a far cleaner signal of local housing markets. Second, MSAs are defined around a central city and its economically integrated hinterland, making them a more suitable unit for studying homelessness as an urban problem. While the Office of Management and Budget (OMB) occasionally updates MSA boundaries, these changes are much more subtle compared to CoCs (MSAs very rarely merge with other MSAs), allowing them to be compared across time with less concern. I also conduct robustness checks to verify that MSA boundary updates do not drive the results.\\
\\
\newpage

\textbf{\large Population Weighted Areal Interpolation}
\normalsize\\

Figure 3 maps the Continuum of Cares (CoCs) operating in and around the St. Louis and Springfield MSAs. As is immediately apparent, the two systems have incongruent boundaries. Five “central-city” CoCs lie entirely within the St. Louis MSA, so their PIT counts can be summed directly. By contrast, the St. Louis MSA overlaps three other CoCs—South Central Illinois, Southern Illinois, and Missouri Balance of State—so those CoC totals must be split before producing an MSA total.

To allocate PIT counts from CoCs to MSAs, I use \textbf{population-weighted areal interpolation:}
\begin{enumerate}
  \item \textbf{Overlay the Inputs.} Overlay CoC polygons with census block
        groups (BGs) as points and obtain each BG’s population
        \(\text{Pop}_{g}\) for every BG \(g\).
  \item \textbf{Disaggregate CoC totals to BGs (mass–preserving).}
        For each CoC \(c\) with PIT total \(H_{c}\), allocate that total across
        its BGs in proportion to population:
        \[
          w_{gc}=\frac{\text{Pop}_{g}}{\sum_{g'\in c}\text{Pop}_{g'}},
          \qquad
          h_{g}=w_{gc}\,H_{c},
        \]
        so that \(\sum_{g\in c} h_{g}=H_{c}\).
  \item \textbf{Aggregate to the MSA.} Sum the BG allocations that fall inside
        MSA \(m\) to obtain the MSA-level count:
        \[
          H_{m}=\sum_{g\in m} h_{g}.
        \]
\end{enumerate}

BGs are used as the “building blocks” because they are the smallest geography for which total population data is available from the Census Bureau. In the 2010 Census there were 217{,}740 BGs, each typically containing 600–3{,}000 residents. This allows for a much finer interpolation than using census tracts, which contains an average of 4000 people per tract.\footnote{In the 2010 US Census, there were 74,134 census tracts in total.}

Figure 5 zooms in on the city-center area (red box in Figure 3), with each point representing a census block group. Yellow squares are BGs inside the St. Louis City CoC,  pink triangles are BGs in the Madison County CoC, and so on. Notice how the BGs are more densely populated in the city center compared to its surrounding regions. Figure 4 illustrates how the interpolation is done by overlaying CoC boundaries, MSA boundaries, and BGs. The CoC-level PIT total is first disaggregated to BGs and then re-aggregated to MSAs. Counts assigned to block groups outside any MSA are excluded, since the analysis focuses on urban homelessness.

\newpage

\newgeometry{left=0.1cm, right=0.1cm, top=1cm, bottom=1cm}

\begin{figure}[H]
    \centering
    \begin{minipage}{0.5\textwidth}
        \centering
        \includegraphics[width=\textwidth]{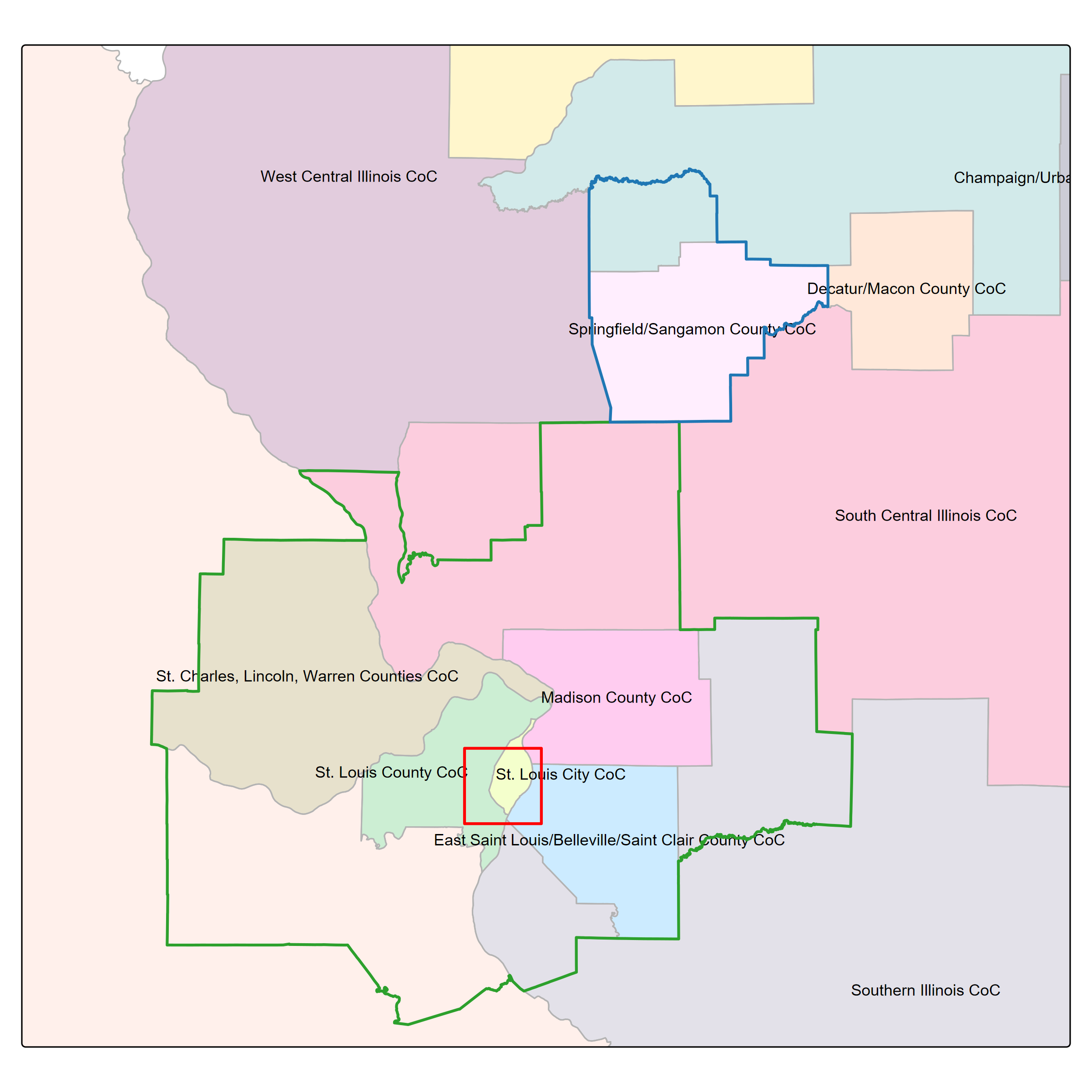}
        \caption{CoCs Operating in St. Louis and Springfield}
        \label{fig:cocs}
    \end{minipage}\hfill
    \begin{minipage}{0.5\textwidth}
        \centering
        \includegraphics[width=\textwidth]{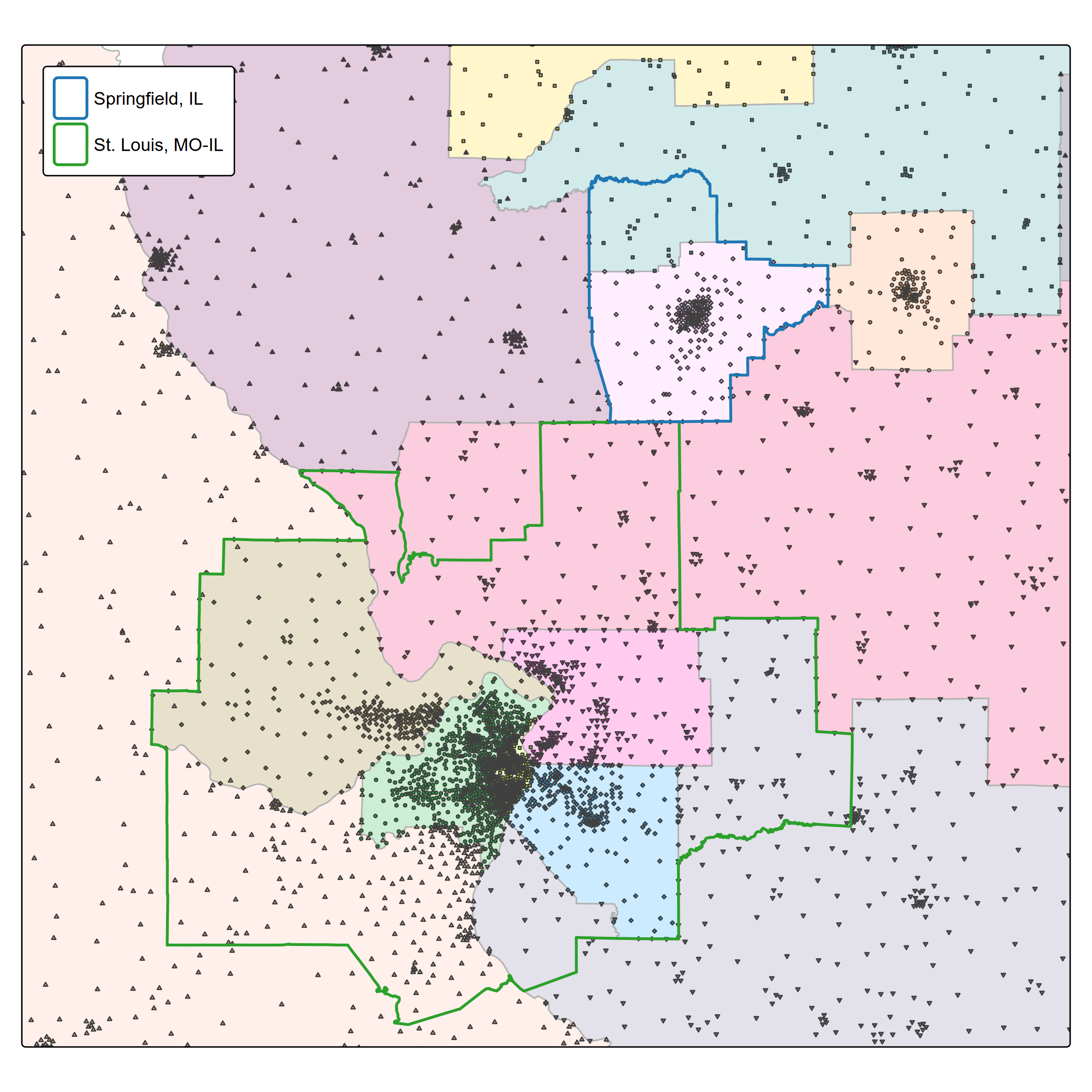}
        \caption{Areal Interpolation from CoCs to MSAs using Block Group \\
        Population as Weights}
        \label{fig:interpolation}
    \end{minipage}
\end{figure}

\begin{figure}[H]
    \centering
    \includegraphics[width=0.5\textwidth]{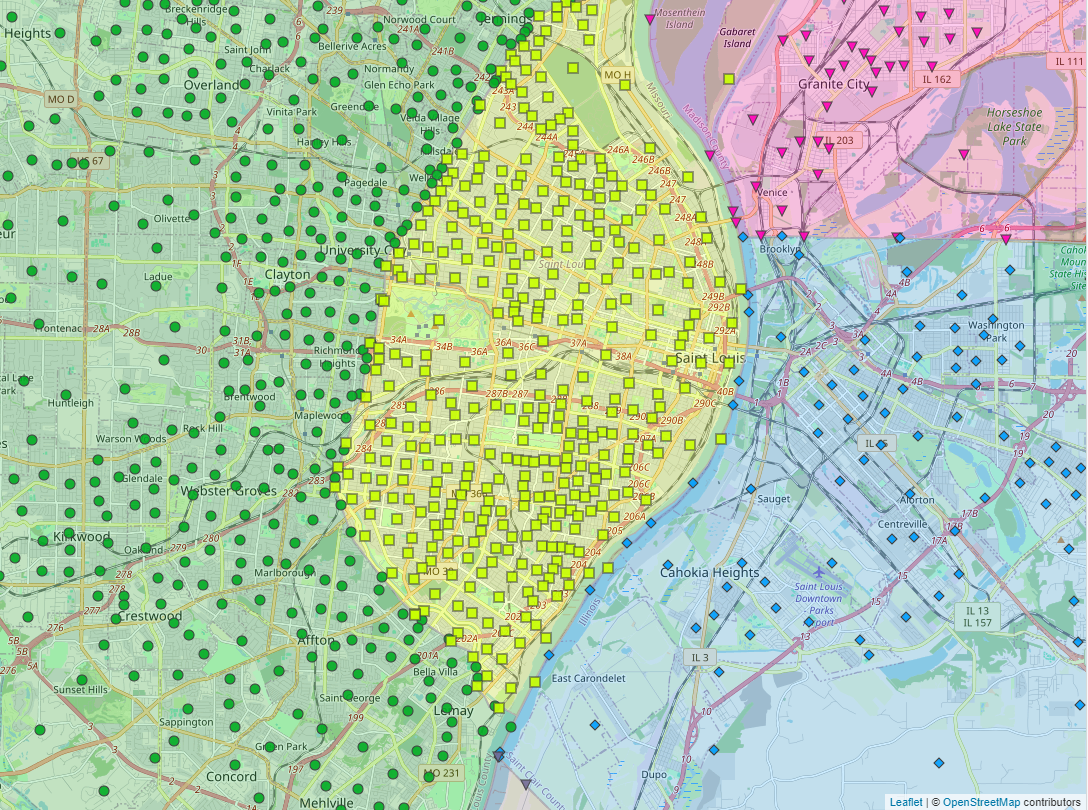}
    \caption{Census Block Groups}
    \label{fig:example}
\end{figure}
\restoregeometry

\newpage
\textbf{\large Data Sanity Check}
\normalsize\\
To ensure that our MSA-level homelessness counts are reliable, I ran a simple validation exercise using data on crowding from the ACS. In its 'Occupancy Characteristics' table, the Census Bureau has data of how many housing units are crowded, meaning units with more than 1.5 occupants per room. Dividing this by the total number of housing units, I computed the share of crowded housing units for each MSA. The reasoning is that since crowdedness is another form of a precarious housing condition, we would expect chronic homelessness rates and crowded rates to be positively correlated. Table 1 reports the simple correlations—both in levels and in logs—between chronic homelessness rates and the crowded rates for each year in my sample. Given that crowding is a housing unit-based statistic while chronic homelessness is population-based, a raw correlation of 0.3 to 0.5 indicate a strong relationship between the two variables. These results provide reassuring evidence of the reliability of my dataset.

\begin{table}[H]
\centering

\begin{tabular}{rrcc}
  \hline
 & Year & Raw Correlation & Correlation of Logged Values \\ 
  \hline
1 & 2011 & 0.300*** & 0.387*** \\ 
  2 & 2016 & 0.433*** & 0.356*** \\ 
  3 & 2020 & 0.526*** & 0.430*** \\ 
   \hline
\end{tabular}
\caption{Correlation Tests Between Crowding and Chronic Homelessness} 
\end{table}

\textbf{\large Summary Statistics}
\normalsize\\

\begin{table}[H]
\centering
\begin{tabular}{rrrrrrr}
  \hline
 & year & count & Mean Crowded Rate & Mean Chronic Rate& Mean Log crowded & Mean Log Chronic \\ 
  \hline
1 &          2011 &           383 & 0.723759791123 & 0.000272397996 & -0.645889788975 & -8.703544381744 \\ 
  2 &          2016 &           384 & 0.799243980128 & 0.000224255776 & -0.496005012251 & -8.944818756740 \\ 
  3 &          2020 &           388 & 0.833699121898 & 0.000266578179 & -0.462424293754 & -8.835314823804 \\ 
   \hline
   \hline
\end{tabular}
\caption{Mean Rates of Crowding and Chronic Homelessness in MSAs by Year} 
\end{table}

\begin{figure}[H]
    \centering
    \includegraphics[width=0.6\textwidth]{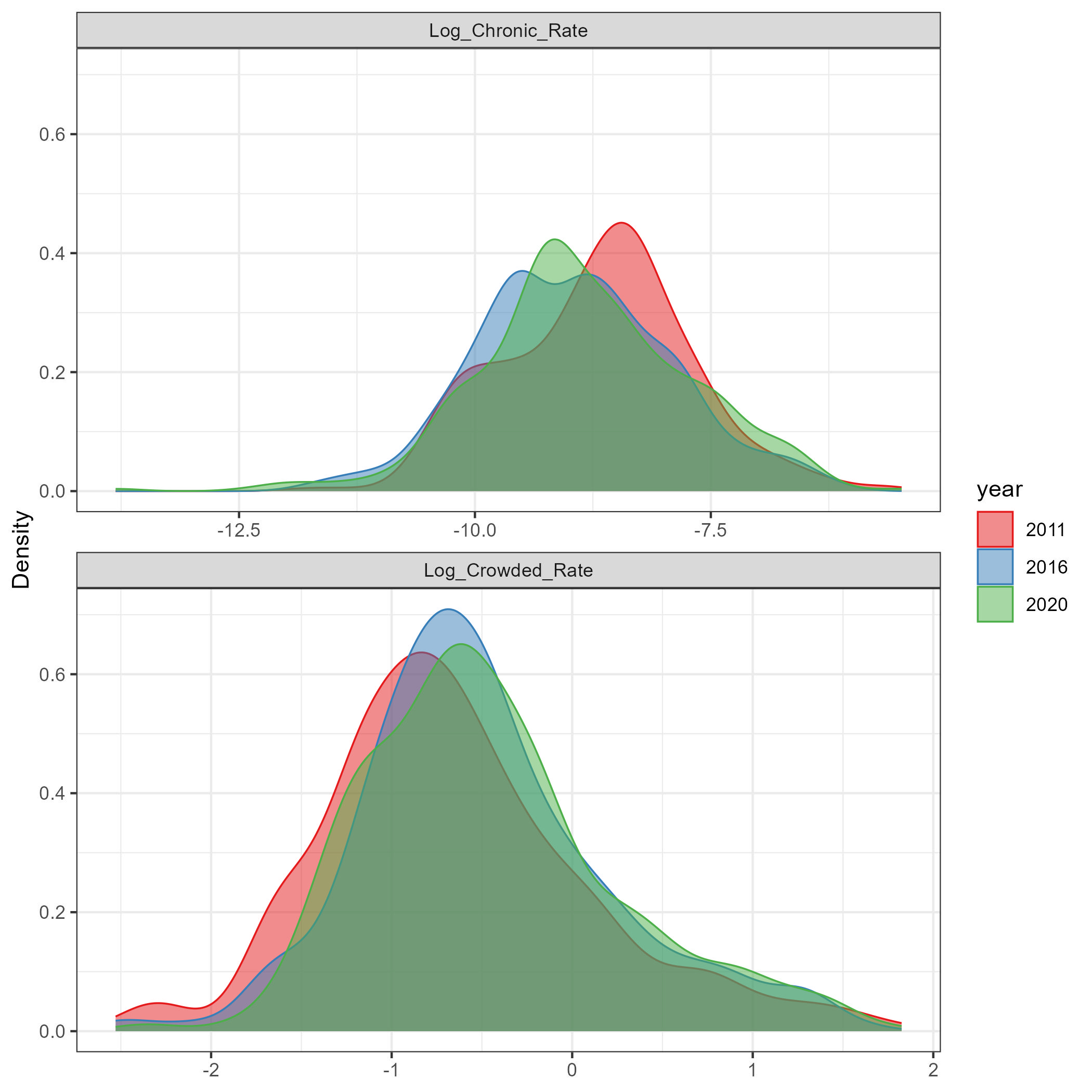}
    \caption{Crowded Rate and Chronic Rate densities by year}
    \label{fig:example}
\end{figure}

\section{Model and Results}

Inspired by Glaeser and Gyourko (2005), I regress the change in chronic homelessness rates and crowded rates of MSAs on a transformation of its rent price changes as shown in equation (1) and (2) below. Changes in rent is entered in piecewise linear form to allow for differential effects in rent increases and decreases. Thus the $Rent Plus_{it}$ variable takes on a value of zero if MSA $i$’s rent decreased during time period $t$ and equals MSA $i$'s actual change in rent if the city saw an increase in rent during the relevant time period. Analogously, the $Rent Minus_{it}$ variable\footnote{Out of the 765 MSA by year observations my sample, 49 of the observations had a decrease in real rent.} equals zero if the MSA saw a rent increase during time period t and equals the change in rent if the MSA's rent fell. Since I pool observations across different time periods (2011-2016 and 2016-2020), I include time dummies $\delta_{t}$ to allow for different intercepts across decades. Standard errors are clustered at the MSA level to correct for intertemporal correlation in the error terms associated with multiple observations of the same MSA over time. All dollar values, including rent, are real values. (in 1999 dollars)

I applied a natural logarithmic transformation to each outcome variable and the unemployment rate due to their highly skewed natures. Rent is also the logged value of rent because my main interest is in estimating the elasticities. As control variables, I include the percentage of population with Public Assistance and the median household income. The share of the population with public assistance is included to capture the fraction of those who are in abject poverty, as public assistance programs such as general assistance and Temporary Assistance to Needy Families (TANF) are strictly means tested. I included other covariates used by previous studies as robustness checks. Including other income quintiles or different poverty measures (SSI recepients, SNAP receipients) did not qualitatively change my results.

\subsection{\large Rent}  
\textbf{\large{OLS}}

\begin{equation}
\Delta y_{it} =\alpha_{0} + \alpha_{1} \Delta Rent Plus_{it} + \alpha_{2} \Delta Rent Minus_{it}+ \alpha_{3}\delta_{t}+ Covariates+\epsilon_{it}
\end{equation}
The first column of table 3 reports results from a specification that pools the 765 observations on changes in log rents and log chronic homelessness rates\footnote{3 observations were lost due to infinite log values} that we have for our sample of MSAs. Among MSAs that saw an increase in rent, the estimate of the coefficient on $\Delta$ log Rent (+) indicates an elasticity of the chronic homelessness rate with respect to rent price changes of 1.74. Thus a one-percent increase in median rent prices is associated with nearly a two-percent increase in chronic homelessness rates in the same period. Among MSAs that saw a decrease in rent, the elasticity of the chronic homelessness rate with respect to rent price change is statistically indistinguishable from 0. The second column presents results for the crowded rate of housing units, showing a similar pattern. Columns 3 and 4 report that there is no economically meaningful change in this pattern if the specification is estimated using longer-run changes between 2011 and 2020. However, the down-side elasticity of rent is more precisely estimated in the models that use the 9-year differences, possibly due to reduced noise in the data. The result in column 3 shows that a 1\% increase in median rent is associated with a 2.1\% increase in chronic homelessness rates, but a decrease in the same amount is associated with a mere 0.04\% decrease in chronic homelessness rates, which is a 52-fold difference in magnitude. As the chi-squared test results reported in table 4 shows, these effects are statistically different from one another in all 4 specifications.

\begin{table}[H]
\caption{OLS Regression Results on Housing Crowding}
\centering
\begin{tabular}{lccccc}
\toprule
& $\Delta$ log Chronic Rate & $\Delta$ log Crowded Rate & $\Delta$ log Chronic Rate & $\Delta$ log Crowded Rate \\ \midrule 
$\Delta$ log Rent (+) & \num{1.740}** & \num{0.986}** & \num{2.100}** & \num{0.922}*\\
& (\num{0.613}) & (\num{0.381}) & (\num{0.672}) & (\num{0.389})\\
$\Delta$ log Rent (–) & \num{-0.009} & \num{0.000} & \num{0.039}* & \num{0.045}***\\
& (\num{0.006}) & (\num{0.005}) & (\num{0.020}) & (\num{0.013})\\
$\Delta$ \%pop with P. A. & \num{0.033}* & \num{0.008} & \num{-0.063} & \num{-0.027}\\
& (\num{0.016}) & (\num{0.014}) & (\num{0.057}) & (\num{0.029})\\
$\Delta$ Median HH Inc. & \num{-0.000}** & \num{0.000} & \num{-0.374} & \num{-0.017}\\
& (\num{0.000}) & (\num{0.000}) & (\num{0.630}) & (\num{0.455})\\
Num.Obs. & \num{765} & \num{768} & \num{383} & \num{384} \\
R2 & \num{0.066} & \num{0.023} & \num{0.031} & \num{0.024} \\
R2 Adj. & \num{0.060} & \num{0.017} & \num{0.021} & \num{0.014}\\
RMSE & \num{0.77} & \num{0.48} & \num{0.95} & \num{0.54} \\
Std.Errors & by: GEOID & by: GEOID & by: GEOID & by: GEOID \\
FE: year & X & X & X & X\\

\bottomrule
\end{tabular}
\end{table} 
Note: 3 observations in column (1) and 1 observation in column (3) were lost to -Inf log Chronic Rate values

\begin{table}[H]
\caption{Chi-Squared Test Results}
\centering
\begin{tabular}{lrrrr}
  \hline
 & Res.Df & Df & Chisq & Pr($>$Chisq) \\ 
  \hline
col (1)  & 760 &  &  & \\ 
  & 759 & 1 & 8.10 & 0.0044\\ 
   \hline

col (2)  & 763 &  &  &  \\ 
    & 762 & 1 & 6.67 & 0.0098 \\ 
  \hline
col (3) & 379 &  &  &  \\ 
   & 378 & 1 & 9.16 & 0.0025 \\ 
   \hline

col (4) & 380 &  &  &  \\ 
   & 379 & 1 & 5.04 & 0.0247 \\    

   \hline

\end{tabular}
\end{table}

\textbf{\large{Method of Moments (Multiplicative Mean)}}

As a robustness check, I model the relationship between homelessness rates and rent in a multiplicative mean specification and use the quasi-differencing moment condition. Here,  $g(X_{i,t}^{'}\beta)$ is $E[y_{it}|X_{it}] = c_{i} exp(X_{i,t}^{'}\beta)$, the conditional mean of MSA $i$'s homelessness rate at time $t$. $c_{i}$ is the time-invariant MSA fixed effect, which gets canceled out in the moment condition. Similarly, changes in rent enter \textbf{the exponent} in piecewise linear form: 

$$\frac{g(X_{i,t}^{'}\beta)}{g(X_{i,t-1}^{'}\beta)}=exp(\alpha_{0} + \alpha_{1} \Delta Rent Plus_{it} + \alpha_{2} \Delta Rent Minus_{it}+ \alpha_{3}\delta_{t}+ Covariates+\epsilon_{it})$$

\begin{equation}
E[y_{i,t} - \frac{g(X_{i,t}^{'}\beta)}{g(X_{i,t-1}^{'}\beta)}y_{i,t-1}] = \textbf{0}
\end{equation}

The results presented in Table 5 tell us a similar story, but this time, the coefficients have a semi-elasticity interpretation. Among MSAs that saw an increase in rent, the estimate of the coefficient on $\Delta$ log Rent (+) indicates a semi-elasticity of chronic homelessness rate with respect to rent price change of 0.001. Thus a \$100 increase in real median rent prices is associated with a 0.14\% increase in chronic homelessness rates in the same period. Among MSAs that saw a decrease in rent, the semi-elasticity of chronic homelessness rate with respect to rent price change is statistically indistinguishable from 0, although it is much less precisely estimated as compared to the additive mean specification from the preceding section. The second column presents results for the crowded rate of housing units, showing a similar pattern.


\begin{table}[htbp]\centering
\def\sym#1{\ifmmode^{#1}\else\(^{#1}\)\fi}
\caption{Method of Moments Estimates}
\begin{tabular}{l*{2}{c}}
\toprule
                    &\multicolumn{1}{c}{(1)}&\multicolumn{1}{c}{(2)}\\
                    &\multicolumn{1}{c}{$\Delta$ Chronic Rate}&\multicolumn{1}{c}{$\Delta$ Crowded Rate}\\
\midrule
 Years 2011-2016 Trend             &     -0.2606\sym{**}&      0.0711\sym{+}  \\
                    &    (0.0599)         &    (0.0391)         \\
\midrule
Years 2016-2020 Trend             &     -0.0803         &     -0.0185         \\
                    &    (0.1112)         &    (0.0480)         \\
\midrule
$\Delta$ Rent (+)               &      0.0014\sym{*} &     0.0007\sym{+}  \\
                    &    (0.0007)         &    (0.0004)         \\
\midrule
$\Delta$ Rent (-)            &      0.0102         &     -0.0018         \\
                    &    (0.0154)         &    (0.0045)         \\
\midrule
$\Delta$ \%pop with Cash Public Assistance            &      0.0355\sym{+}  &     -0.0017         \\
                    &    (0.0212)         &    (0.0151)         \\
\midrule
$\Delta$ Median Household Income             &     -0.0000\sym{**}&      0.0000         \\
                    &    (0.0000)         &    (0.0000)         \\
\midrule
Obs.                &         765         &         768         \\
Std.Errors clustered by GEOID            &                     &                     \\
\bottomrule
\multicolumn{3}{l}{\footnotesize Standard errors in parentheses}\\
\multicolumn{3}{l}{\footnotesize \sym{+} \(p<.10\), \sym{*} \(p<.05\), \sym{**} \(p<.01\)}\\
\end{tabular}
\end{table}

\textbf{\large Robustness Checks: Other Income Quintiles and Rent Percentiles}

Here, I present robustness check results for different household income quintiles and different rent percentiles. My main results are robust to replacing the median household income with different income quintiles. Still, they are not fully robust to using different rent percentiles instead of the median rent. 

However, it should not be concluded that my main results were a fluke. While the median rents of MSAs were readily available from the Census Bureau as the gross rent variable 'B25031,' rents at percentiles other than the median were not. Therefore, they had to be imputed by taking the percentiles of rent using the microdata samples I obtained from IPUMS, a process that likely introduced a lot of noise. Since the smallest unit of geography identifiable in the PUMS is the PUMA, I linked PUMAs to MSAs depending on whether the majority of the PUMA population resided in the MSA boundary. The resulting match was a very fuzzy one, as there were many PUMA-MSA matches where the fraction of PUMA population in the MSA barely exceeded the threshold (e.g., 51\%).\footnote{MSAs are areas with more than 50,000 population. PUMA population has to exceed 10,000 to preserve anonymity of the ACS survey repondents.} Then, I calculated the rent percentiles in MSAs using all people who resided in PUMAs matched with the MSA. Therefore, in many cases, those who live outside the MSA boundaries were included in the calculation of rent percentiles, which would never have happened with the Census Bureau's median rent. The fact that the OMB adopted new PUMA boundaries every decade introduces more potential for noise, as the PUMA-MSA links are not consistent across years. The fact that the coefficients are much less precisely estimated when we substitute the Census Bureau median rent with our PUMS-reconstructed median rent lends support to the interpretation that noise is driving coefficients to be imprecisely estimated and therefore untrustworthy. I am currently in the process of looking for more credible data sources for rent at the MSA level. HUD's Fair Market Rents seem to be a possible candidate, but the fact that the HUD changed their methodology for calculating FMRs during my study period introduces complications.

\newpage

\textbf{Chronic Homelessness Rate}
\begin{table}[H]
\centering
\begin{tabular}{lccccc}
\toprule

& (1) & (2) & (3) & (4) & (5) \\ \midrule 
$\Delta$ log Rent (+) & \num{1.785}** & \num{1.723}* & \num{2.007}** & \num{2.164}** & \num{1.629}* \\
& (\num{0.664}) & (\num{0.735}) & (\num{0.756}) & (\num{0.729}) & (\num{0.671}) \\
$\Delta$ log Rent (–) & \num{-0.009} & \num{-0.009} & \num{-0.008} & \num{-0.008} & \num{-0.009} \\
& (\num{0.006}) & (\num{0.006}) & (\num{0.006}) & (\num{0.006}) & (\num{0.006}) \\
$\Delta$ \%pop with P.A. & \num{0.034}* & \num{0.033}* & \num{0.031}+ & \num{0.029}+ & \num{0.034}* \\
& (\num{0.017}) & (\num{0.016}) & (\num{0.017}) & (\num{0.017}) & (\num{0.016}) \\
$\Delta$ log 1st Quintile Inc. & \num{-0.091} &  &  &  &  \\
& (\num{0.330}) &  &  &  &  \\
$\Delta$ log 2nd Quintile Inc. &  & \num{0.024} &  &  &  \\
&  & (\num{0.628}) &  &  &  \\
$\Delta$ log 3rd Quintile Inc. &  &  & \num{-0.584} &  &  \\
&  &  & (\num{0.790}) &  &  \\
$\Delta$ log 4th Quintile Inc. &  &  &  & \num{-1.021} &  \\
&  &  &  & (\num{0.876}) &  \\
$\Delta$ log 5th Quintile Inc. &  &  &  &  & \num{0.272} \\
&  &  &  &  & (\num{0.615}) \\
Num.Obs. & \num{765} & \num{765} & \num{765} & \num{765} & \num{765} \\
R2 & \num{0.064} & \num{0.064} & \num{0.064} & \num{0.065} & \num{0.064} \\
R2 Adj. & \num{0.058} & \num{0.058} & \num{0.058} & \num{0.059} & \num{0.058} \\
RMSE & \num{0.77} & \num{0.77} & \num{0.77} & \num{0.77} & \num{0.77} \\
Std.Errors & by: GEOID & by: GEOID & by: GEOID & by: GEOID & by: GEOID \\
FE: year & X & X & X & X & X \\

\bottomrule
\end{tabular}
\end{table}

\begin{table}[H]
\centering
\begin{tabular}{lcccc}
\toprule
& (1) & (2)  & (3)   & (4)    \\ \midrule 
$\Delta$ Rent 5th pct (IPUMS) (+) & \num{1.571}** &  &  &  \\
& (\num{0.525}) &  &  &  \\
$\Delta$ Rent 5th pct (IPUMS) (-) & \num{-1.196}+ &  &  &  \\
& (\num{0.644}) &  &  &  \\
$\Delta$ Rent 15th pct (IPUMS) (+) &  & \num{-0.225} &  &  \\
&  & (\num{1.195}) &  &  \\
$\Delta$ Rent 15th pct (IPUMS) (-) &  & \num{1.615} &  &  \\
&  & (\num{1.199}) &  &  \\
$\Delta$ Rent 25th pct (IPUMS) (+) &  &  & \num{0.992} &  \\
&  &  & (\num{1.870}) &  \\
$\Delta$ Rent 25th pct (IPUMS) (-) &  &  & \num{2.283} &  \\
&  &  & (\num{1.434}) &  \\
$\Delta$ Rent 50th pct (IPUMS) (+) &  &  &  & \num{4.257}* \\
&  &  &  & (\num{1.979}) \\
$\Delta$ Rent 50th pct (IPUMS) (–) &  &  &  & \num{1.337} \\
&  &  &  & (\num{1.626}) \\

$\Delta$ \%pop with P.A. & \num{-0.070} & \num{-0.070} & \num{-0.072} & \num{-0.076} \\
& (\num{0.054}) & (\num{0.056}) & (\num{0.056}) & (\num{0.055}) \\
$\Delta$ log Median Household Income & \num{0.239} & \num{0.126} & \num{0.056} & \num{-0.029} \\
& (\num{0.575}) & (\num{0.586}) & (\num{0.588}) & (\num{0.589}) \\

Num.Obs. & \num{377} & \num{377} & \num{377} & \num{377} \\
R2 & \num{0.023} & \num{0.009} & \num{0.014} & \num{0.020} \\
R2 Adj. & \num{0.012} & \num{-0.002} & \num{0.004} & \num{0.009} \\
RMSE & \num{0.96} & \num{0.96} & \num{0.96} & \num{0.96} \\
Std.Errors & by: GEOID & by: GEOID & by: GEOID & by: GEOID \\
FE: year & X & X & X & X \\

\bottomrule
\end{tabular}
\end{table}

\section{An Economic Model of Sticky Homelessness}
This model builds on O'Flaherty (1995) to explain the asymmetric responses of homelessness to changes in rent that I identified in the data. The key intuition is that being homeless in time $t$ has a negative effect on a person's future income in time $t+1$, which in turn serves to perpetuate his or her homeless spell. Modeling the exact mechanism through which income is affected by lack of housing is a work in progress, with potential ideas of modifying the job search model with parameters that are endogenous to homelessness status. 
\\
\\
\textbf{\large The Representative Consumer and the Homeless Person}

Suppose the consumer consumes two goods: housing, denoted as $H$, and a numeraire good, denoted as $x$. A larger $H$ means a housing unit of higher quality. Then, the consumer's utility function can be represented as $U(H, Y-P (H)),$ where $P(H)$ is the price of housing of quality $H$. The consumer solves the utility maximization problem:
\begin{equation}
\max_{H,x} U(H,x) \quad \text{s.t. } P(H)H + x = Y
\end{equation}
A person is homeless if she maximizes her utility at $H=0$.\\
Additional assumptions imposed on the consumer's utility function are: \\
1) $u(.,.)$ is strictly increasing in both arguments.\\
2) $u(.,.)$ is is twice continuously differentiable in $x$ with $U_{xx}<0$\\
3) $H$ and $x$ are compliments. In other words, $H_{1} \leq H_{2}$ implies $\frac{\partial U(H_{1},x)}{\partial x}\leq \frac{\partial U(H_{2},x)}{\partial x}$.\\
Also assume that $P(H)$ is increasing in $H$. Better quality housing is more expensive.
\\
\\
\textbf{\large The Homeless Bid-Rent Curve}\\
For each consumer, the 'homeless bid-rent curve,' $B^{H}(H,Y),$ can be defined. That is, a consumer is indifferent between being homeless and consuming quality $H$ housing for price $B^{H}(H,Y).$
$$U(H,Y-B^{H}(H,Y)) = U(0,Y)$$
The homeless bid-rent curve is an increasing function of $Y.$ Rich people are willing to pay more money to avoid becoming homeless. \\

\[
\frac{\partial U(0,Y)}{\partial x}
= \frac{\partial U(H,\, Y - B^{H}(H,Y))}{\partial x}
\left[\,1 - \frac{\partial B^{H}(H,Y)}{\partial Y}\,\right].
\]

\noindent From assumptions $2)$ and $3)$,
\[
0 \;<\; \frac{\partial U(0,Y)}{\partial x}
\;\le\; \frac{\partial U(H,Y)}{\partial x}
\;<\; \frac{\partial U(H,\, Y - B^{H}(H,Y))}{\partial x}.
\]

\noindent Thus
\begin{equation}
    0 \;<\; \frac{\partial B^{H}(H,Y)}{\partial Y} \;<\; 1.
\end{equation}
\textbf{\large The Cutoff Income}\\
From this, it follows that there exists $\bar{Y}$ such that $B(H,\bar{Y}) = P(H_{min}).$ $\bar{Y}$ is the 'cutoff income,' which is the amount of income below which a person cannot afford the minimum quality housing offered at the private rental market and becomes homeless. There could be cases where a person's utility is maximized at $H=0$ even though there are other qualities of housing in his or her budget set, but we assume away such cases at the moment. Therefore, whoever has income less than $\bar{Y}$ is homeless, and whoever has a higher income is not. 
\\
\\
\textbf{\large Dynamic Effects of Being Homeless}\\
I assume that being homeless now is detrimental to an agent's income in the future. A person who is homeless in time $t$ receives a negative shock $\delta<0$ to his or her income in the next period $t+1$.
\begin{equation}
Y_{t+1}=Y_{t}+\delta\cdot\mathbf{1}({H_{t}=0}) + \epsilon_{t+1}
\end{equation}
A large literature documents that people experiencing homelessness face substantially higher burdens of chronic disease, mental illness, and substance use—and suffer mortality rates several times those of the general population. Brown, R., et al. (2019) cite evidence that homeless individuals often cannot store medications properly, maintain a healthy diet, or access regular check-ups, leading to a worsening of pre-existing conditions like diabetes, hypertension, and asthma. These adverse health shocks have well-established causal effects of lowering subsequent employment and earnings. García-Gómez, et. al (2014) identify the causal effects of sudden illness, represented by acute hospitalizations, on employment and income. Consistent with this, observed employment probabilities drop sharply during episodes of homelessness and recover only partially, even with services. Together, these findings support modeling homelessness at time $t$ as imposing a negative shock to next-period income $Y_{t+1}$.\footnote{I also have anecdotal evidence from my work as a volunteer. Many of the homeless told me that it is nearly impossible to apply for most jobs because they do not have a fixed address and cannot access banking services properly. Stories of people getting evicted and losing their jobs as a result of the eviction are well-documented in the book \textit{Evicted} as well.} Modeling exactly how deterioration in health would affect an agent's future income is a work in progress, building on Grossman (1972)'s model of health production and job/housing search models.\\


\textbf{\large The Low-Quality Housing Market}

Those who enter the lowest-quality housing market are 1) those who maximize their utility by consuming $H_{min},$ and 2) those who are homeless. The bid-rent for $H_{min}$ of the 1) type is the amount of money they are willing to give up that makes them just indifferent between consuming their next preferred quality of housing, $H^{**}$. This is denoted as $B^{1}(H_{min},Y)$ in equation (6).
\begin{equation}
U(H_{min},Y-B^{1}(H_{min},Y)) = U(H^{**},Y-P(H^{**}))
\end{equation}

The homeless can be thought of market participants just with very low willingness to pay for housing due to their small income. In other words, the homelessness have $B^{H}(0,Y)<P(H_{min})$, where $B^{H}(0,Y)$ is the homeless bid-rent defined in the previous subsection. Then, we can construct the housing demand curve for $H_{min},$ which is the collection of bid-rents for $H_{min}$ of every participants in the $H_{min}$ quality of housing. Those whose bid-rent is less than $P(H_{min})$ are homeless, as they cannot afford the lowest quality housing unit supplied in the housing market. (The right side of the demand curve in Figure 7). Those who have bid-rents higher than $P(H_{min})$ consume a unit of $H_{min}$ from the market, and they are housed (The left side of the demand curve in Figure 7). It should be noted that a higher income does not necessarily mean a larger $B^{1}$, as $B^{1}$ is defined only for consumers whose utility-maximizing choice of housing is $H_{min}$ in the first place.

Using the $H_{min}$ demand curve I constructed, I present an explanation for the asymmetry observed in my results. The homeless in period $t$ have a bid-rent of $B^{H}(0,Y_{t}).$ However, following equations (4) and (5), their bid-rent decreases to $B^{H}(0,Y_{t+1})$ in period $t+1$ because $Y_{t}>Y_{t+1}$. From the market’s perspective this is a right-hand inward pivot of the demand curve, with the left segment unchanged. As a result, a kink appears at the initial equilibrium at $t$: the curve follows the original demand up to that point and then continues to the right with a sharply downward segment. This dynamic is illustrated in Figure 8.

\begin{figure}[H]
    \centering
    \includegraphics[width=0.5\textwidth]{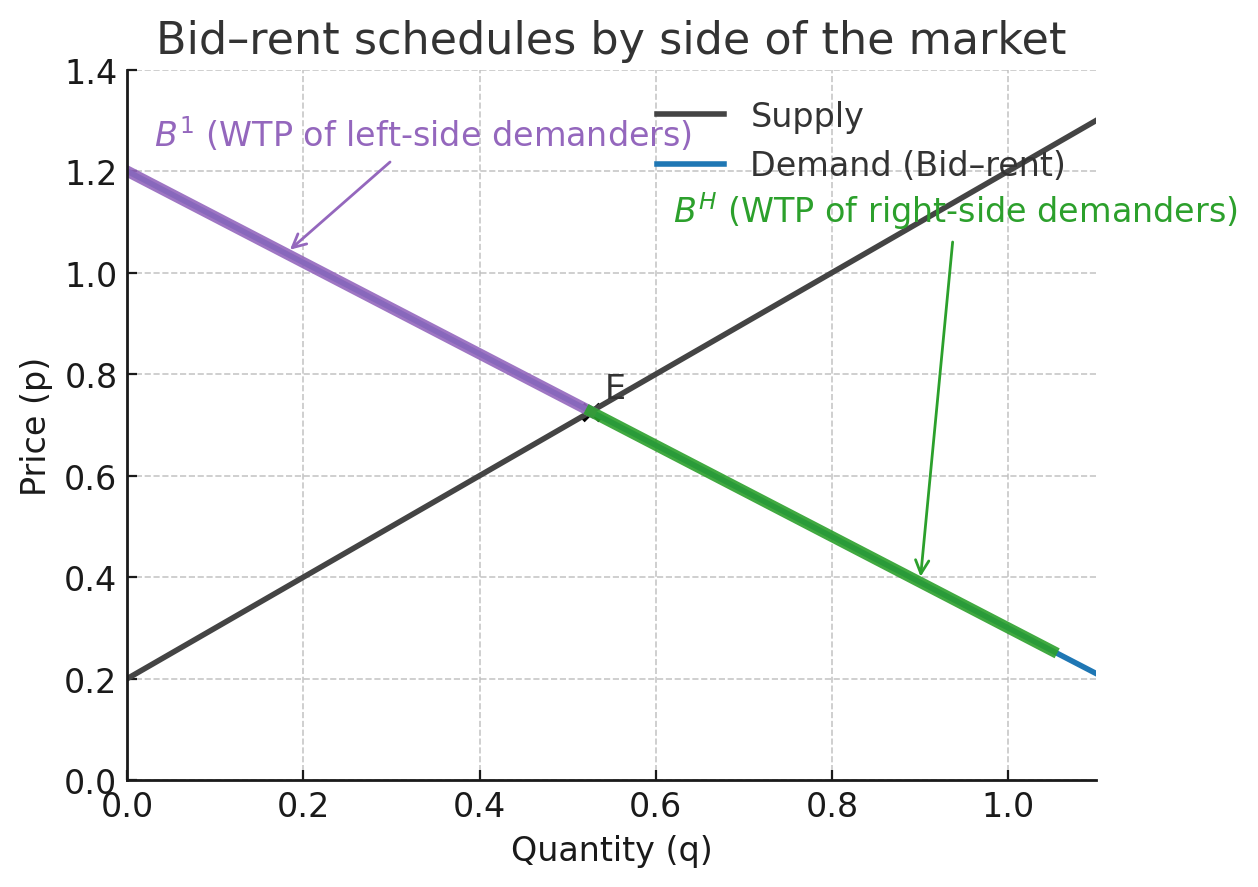}
    \caption{The $H_{min}$ Market}
    \label{fig:example}
\end{figure}

\begin{figure}[H]
    \centering
    \includegraphics[width=1\textwidth]{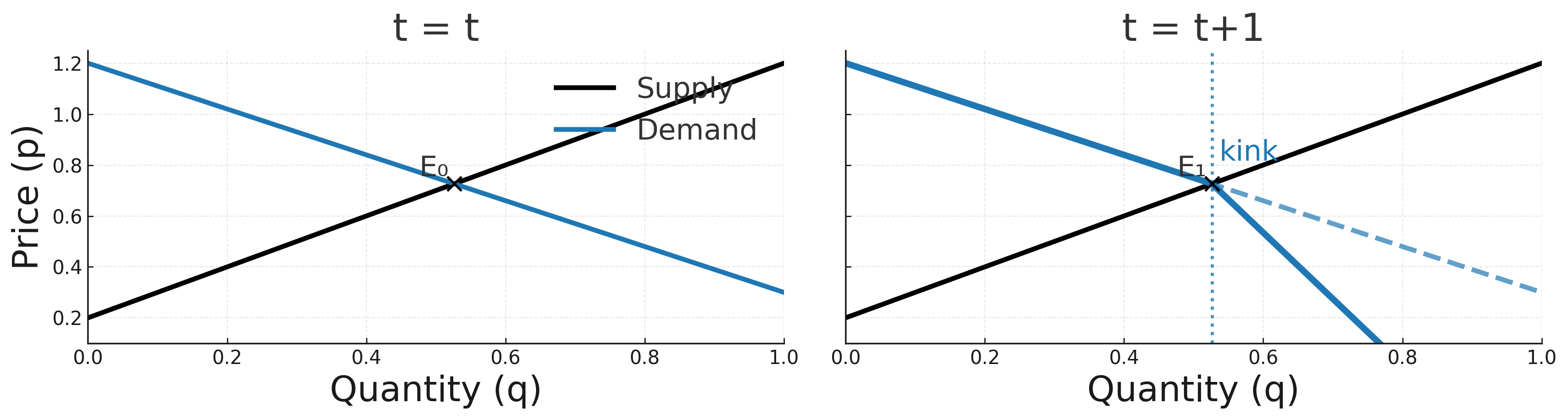}
    \caption{The Market at periods $t$ and $t+1$}
    \label{fig:example}
\end{figure}

With this demand curve, consider what we would observe when there is a supply shock in the market between period $t$ and $t+1$. To a negative shock in supply, we would observe an increase in price and a decrease in the quantity of $H_{min}$ housing units consumed. In other words, we would observe more homelessness, as a sizeable portion of the people who got priced out of $H_{min}$ will end up homeless in period $t+1$. On the other hand, to a positive supply shock, we would observe a decrease in prices but almost no change in the quantity. These predictions are consistent with my regression results that were presented in the earlier sections. To show that supply shifts actually dominated in my study period of 2011 to 2021, I regressed the log of vacancy rates on median rent for the MSAs in our sample. The results presented in table 8 clearly show that price and quantity changes were driven mostly by supply shocks, not demand shocks.

\begin{table}[H]
\caption{Positive Relationships between Median Rent and Vacancy Rates}
\centering
\begin{tabular}{lcc}
\toprule

& $\Delta$ Log Vacancy Rate & $\Delta$ Median Rent\\ \midrule 
$\Delta$ Median Rent & \num{0.001}** &  \\
& (\num{0.000}) &  \\
$\Delta$ log Vacancy Rate &  & \num{12.106}** \\
&  & (\num{4.005}) \\
Num.Obs. & \num{768} & \num{768} \\
R2 & \num{0.010} & \num{0.248} \\
R2 Adj. & \num{0.007} & \num{0.246} \\
RMSE & \num{0.35} & \num{43.87} \\
Std.Errors & by: GEOID & by: GEOID \\
FE: year & X & X \\

\bottomrule
\end{tabular}
\end{table}

\section{Instrumenting for Rent}

To be clear, my coefficients on rent increases and rent decreases in equation (1) do not have a causal interpretation. Renters in MSAs that experience a sharp increase in homelessness could choose to migrate to different MSAs, driving down rent prices. There could be other unobserved forces in the urban economy that affect both rents and homelessness rates. To mitigate these concerns, I instrument for changes in rent using the changes in rent predicted by MSA-specific fluctuations in the labor market. 

Following Blanchard and Katz (1992), I create a Bartik (1991) style shift-share instrument using the MSA industry employment shares and the national employment growth rates for each industry. This is a 
weighted average of the growth rates of national industry employment (aggregated to two-digit NAICS categories) with the weights calculated as 
the previous year share of MSA employment in each industry. The idea behind the Bartik instrument is that as MSAs produce different bundles of goods, they experience different shocks to labor demand and thus experience MSA-specific fluctuations, which can be used as an exogenous source of variation in the local labor market. Then, I add interactions of the Bartik predictor instrument with \textbf{$\boldsymbol{\eta_{i}}$}, a vector of local housing supply elasticity estimates adopted from Saiz (2010), the share of land unavailable for development, and the Wharton Residential Land Use Regulatory Index (Gyourko, et. al 2008). 

The idea to interact the Bartik instrument with $\boldsymbol{\eta_{i}}$ was inspired by Baum-Snow and Han (2024). In the paper, the authors interact housing price with characteristics of census tracts to allow for observed heterogeneity in supply elasticities. In my context, allowing for a heterogeneous first-stage relationship between predicted employment growth and changes in rent will serve to increase the power of my mix instrument by reflecting heterogeneity in the pass-through from labor-demand shocks to rents. MSAs that are more supply-constrained (and thus have low elasticity estimates) respond more dramatically to exogenous employment shocks. The fact that housing prices and rents respond to local labor markets is a well-documented phenomenon. Blanchard and Katz (1992) find that relative housing prices decrease steadily to reach a trough of 2 percent after four to five years in response to a negative employment shock, and then return to their previous level over time. Greenstone, Hornbeck \& Moretti (2010) find that counties that “win” million-dollar plants see sizable increases in property values/housing prices alongside employment and wage gains.

My instruments $\boldsymbol{\ln \tilde{\Delta Rent}}$ is constructed as in equation (7) and (8). In equation (7),  $L_{ik}^{t}$ is the share of industry $k$ employment in MSA $i$ in year $t$ and $E_{k}$ is the national employment in industry $k$. \textbf{$\boldsymbol{\eta_{i}}$} is the vector of estimated supply elasticities and observable supply constraints. Undevelopable land share is constructed using satellite data from the US Geological Survey’s National Elevation Database. To be consistent with Saiz (2010), it is defined as the fraction of area within 50 kilometers of the CBD of each region that is undevelopable because of steep slopes, water, or wetlands. The WRI captures the intensity of local growth control policies in a number
 of dimensions, including zoning regulations or project
 approval practices that constrain new residential real estate development. 

\footnote{The data for the undevelopable land share is using satellite data from the US Geological Survey’s National Elevation Database.}

\begin{equation}
    \Delta \ln Bartik_{i,t+1} =  \left[\left\{\sum_{k} \frac{L_{ik}^{t}}{\sum_k L_{ik}^{t}} ( \ln E_{k}^{t+1} - \ln E_{k}^{t} )\right\} \right],  \boldsymbol{\eta_{i}} =  \begin{pmatrix}
1 \\
WRI_{i} \\
Elasticity_{i} \\
\text{\% undevelopable land in $MSA_{i}$}
\end{pmatrix}
\end{equation}

\begin{equation}
\resizebox{0.35 \linewidth}{!}{$    
    \boldsymbol{\tilde{\Delta \ln Rent_{i,t+1}}} =  \Delta \ln Bartik_{i,t+1} \cdot \boldsymbol{\eta_{i}}$}
\end{equation}

My first stage regressions are presented as in equation (9). The actual change in MSA employment rate (not the Bartik instrument), $\Delta\ln E_{i,t+1}$, is included as a control variable to control for the direct effect of employment rate changes might have on homelessness.\footnote{I will have to change my regression specification in equation (1) and include $\Delta\ln E_{i,t+1}$ as a regressor.} $\Delta\ln E_{i,t+1}$ is treated as an endogenous variable, so there are three endogenous variables in total: $\Delta RentPlus_{i,t+1}$, $\Delta RentMinus_{i,t+1}$, and $\Delta\ln E_{i,t+1}$. With four instruments and three endogenous variables, the regression has more exclusion restrictions than endogenous variables, and all parameters are fully identified.\footnote{There is no particular reason for using subscript $t+1$ instead of $t$.} Since my system is over-identified, I will refit the system four times, each time dropping one instrument as a robustness check. I will also report results of the Stock and Yogo test and the Hansen J test.

\begin{equation}
\begin{pmatrix}
    
\Delta RentPlus_{i,t+1}
 =\pi_{0}^{+} + \boldsymbol{\pi_{1}^{+}} \boldsymbol{\tilde{\Delta \ln Rent_{i,t+1}}} +\pi_{2}^{+}\delta_{t+1}+\pi_{3}^{+}\cdot Covariates \\

 \Delta RentMinus_{i,t+1}
 =\pi_{0}^{-} + \boldsymbol{\pi_{1}^{-}} \boldsymbol{\tilde{\Delta \ln Rent_{i,t+1}}} + \pi_{2}^{-}\delta_{t+1}+\pi_{3}^{-}\cdot Covariates\\

  \Delta E_{i,t+1}
 =\pi_{0} + \boldsymbol{\pi_{1}} \boldsymbol{\tilde{\Delta \ln Rent_{i,t+1}}} +\pi_{2}\delta_{t+1}+\pi_{3}\cdot Covariates

\end{pmatrix}
\end{equation}

My Bartik instrument will satisfy the exogeneity condition as long as the national growth rates are not correlated with labor supply shocks in the MSA. This condition, in turn, will be true as long as a sector is not concentrated in a particular MSA - a condition that is satisfied at the two-digit NAICS level.\footnote{The NAICS codes start broad and become more specific, with the first two digits indicating the economic sector and subsequent digits breaking it down into subsectors, industry groups, and finally, specific industries.} It is hard to imagine supply constraints, such as stringency of building regulations and topography, affecting changes in homelessness through a channel other than rents and housing prices. Hilly places may deter people from living on the streets to some extent, but it should be noted that time-invariant factors get differenced out in my long-differencing specification. Moreover, all of the elements in $\boldsymbol{\eta_{i}}$ will be fixed at a time period earlier than my study period, which is 2011 to 2020. Therefore, I am confident that my instruments generate exogenous variations in rent changes.\\

\textbf{\large Note 1} \normalsize

Instead of using the actual change in employment, $\Delta\ln E_{i,t+1}$, as controls, I am also considering using the predicted change in employment $\Delta \ln \hat{E_{i,t+1}}$ itself. In this case, $\Delta \ln \hat{E_{it+1}}$ should be treated as exogenous with homelessness rates and does not have to be instrumented for. In this case, I would have two endogenous variables: $\Delta RentPlus_{i,t}$ and $\Delta RentMinus_{i,t}$. \\

\textbf{\large Note 2} \normalsize

I ran the same analysis with unemployment rates, but I decided not to pursue it further. The results are presented here. The data comes from the BLS LAUS. \\
\\
\textbf{\large [2] UNEMPLOYMENT}  
\subsection{OLS}

$$\Delta y_{it} =\alpha_{0} + \alpha_{1} \Delta Unemp Plus_{it} + \alpha_{2} \Delta Unemp Minus_{it}+ \alpha_{3}\delta_{t}+ Covariates+\epsilon_{it}$$

\begin{table}[H]
\centering
\begin{tabular}{lccc}
\toprule

& $\Delta$ log Chronic Rate & $\Delta$ log Chronic Rate  & $\Delta$ log Crowded Rate \\ \midrule 
$\Delta$ log Unemp (+) & \num{0.402}+ & \num{0.427}* & \num{0.068} \\
& (\num{0.211}) & (\num{0.211}) & (\num{0.100}) \\
$\Delta$ log Unemp (–) & \num{-0.258} & \num{-0.256} & \num{-0.132} \\
& (\num{0.205}) & (\num{0.205}) & (\num{0.151}) \\
$\Delta$ \%pop with Cash Public Assistance & \num{0.013} & \num{0.013} & \num{-0.016} \\
& (\num{0.024}) & (\num{0.024}) & (\num{0.015}) \\
$\Delta$ Median Household Income & \num{-0.000}* &  & \num{0.000} \\
& (\num{0.000}) &  & (\num{0.000}) \\
Num.Obs. & \num{758} & \num{758} & \num{761} \\
R2 & \num{0.066} & \num{0.065} & \num{0.015} \\
R2 Adj. & \num{0.060} & \num{0.060} & \num{0.009} \\
RMSE & \num{0.77} & \num{0.77} & \num{0.48} \\
Std.Errors & by: GEOID & by: GEOID & by: GEOID \\
FE: year & X & X & X \\

\bottomrule
\end{tabular}
\end{table} 
Note: 7 MSAs in Puerto Rico were dropped because they had missing unemployment rates in year 2020 as a result of pandemic-induced data collection problems.  
\subsection{Chi-Squared Test}

$\Delta$ log Chronic Rate
$$H_{0}: \alpha_{1} = \alpha_{2}$$

\begin{table}[H]
\centering
\begin{tabular}{lrrrr}
  \hline
 & Res.Df & Df & Chisq & Pr($>$Chisq) \\ 
  \hline
1 & 753 &  &  &  \\ 
  2 & 752 & 1 & 5.48 & 0.0192 \\ 
   \hline
\end{tabular}
\end{table}

$\Delta$ log Crowded Rate
$$H_{0}: \alpha_{1} = \alpha_{2}$$

\begin{table}[H]
\centering
\begin{tabular}{lrrrr}
  \hline
 & Res.Df & Df & Chisq & Pr($>$Chisq) \\ 
  \hline
1 & 756 &  &  &  \\ 
  2 & 755 & 1 & 19.57 & 0.0000 \\ 
   \hline
\end{tabular}
\end{table}

\subsection{Method of Moments}
$$\frac{g(X_{i,t}^{'}\beta)}{g(X_{i,t-1}^{'}\beta)}=exp(\alpha_{0} + \alpha_{1} \Delta Unemp Plus_{it} + \alpha_{2} \Delta Unemp Minus_{it}+ \alpha_{3}\delta_{t}+ Covariates+\epsilon_{it})$$\\
and $E[y_{it}|X_{it}] = g(X_{i,t}^{'}\beta) = c_{i} exp(X_{i,t}\beta)$.

\begin{table}[htbp]\centering
\def\sym#1{\ifmmode^{#1}\else\(^{#1}\)\fi}
\caption{Method of Moments Estimates}
\begin{tabular}{l*{2}{c}}
\toprule
                    &\multicolumn{1}{c}{(1)}&\multicolumn{1}{c}{(2)}\\
                    &\multicolumn{1}{c}{$\Delta$ Chronic Rate}&\multicolumn{1}{c}{$\Delta$ Crowded Rate}\\
\midrule
Years 2011-2016 Trend            &     -0.1905         &     -0.2923\sym{+}  \\
                    &    (0.2015)         &    (0.1501)         \\
\midrule
Years 2016-2020 Trend            &     -0.0238         &     -0.1058\sym{+}  \\
                    &    (0.1324)         &    (0.0568)         \\
\midrule
$\Delta$ Unemp (+)               &      0.0129         &     -0.0026         \\
                    &    (0.0337)         &    (0.0147)         \\
\midrule
$\Delta$ Unemp (-)            &     -0.0059         &      0.0168         \\
                    &    (0.0468)         &    (0.0323)         \\
\midrule
$\Delta$ \%pop with Cash Public Assistance            &      0.1908\sym{**}&     -0.0277         \\
                    &    (0.0533)         &    (0.0327)         \\
\midrule
$\Delta$ Median Household Income             &     -0.0000\sym{**}&     -0.0000\sym{**}\\
                    &    (0.0000)         &    (0.0000)         \\
\midrule
Obs.                &         761         &         761         \\
Std.Errors clustered by GEOID            &                     &                     \\
\bottomrule
\multicolumn{3}{l}{\footnotesize Standard errors in parentheses}\\
\multicolumn{3}{l}{\footnotesize \sym{+} \(p<.10\), \sym{**} \(p<.05\), \sym{***} \(p<.01\)}\\
\end{tabular}
\end{table}

\newpage

\nocite{*}
\bibliographystyle{apalike}
\bibliography{references}

\end{document}